\documentclass[12pt,a4paper]{article}
\usepackage{amsmath}
\usepackage{graphicx}
\usepackage{times}
\usepackage{xcolor}
\usepackage{cite}
\usepackage{physics}
\begin{document}
\begin{titlepage}
\title{Centrality and cumulative activities  in hadron collisions}
\author{ S.M. Troshin, N.E. Tyurin\\[1ex]
\small  \it NRC ``Kurchatov Institute''--IHEP\\
\small  \it Protvino, 142281, Russian Federation,\\
\small Sergey.Troshin@ihep.ru
}
\normalsize
\date{}
\maketitle

\begin{abstract}
  \color{blue} The notion of  cumulative activity for elastic events  in hadron collision is introduced. \color{black} 
   The interrelation between centrality and cumulative activities of the elastic and inelastic events has been established for hadron scattering. Considerations on its application are presented.  
\end{abstract}
\end{titlepage}
\setcounter{page}{2}
\section{Introduction. The distinctive features of  reflective scattering mode}
 
An apparently natural  expectation is   that  hadron collision energy growth must be accompanied by an  increase  of the relative weight  of the inelastic interactions  because the hadrons   are  extended objects. However, a significantly increasing relative contribution of the elastic scattering events  to $pp$--interactions was observed \cite{totem}.   

The elastic scattering matrix element  in the impact parameter representation $S(s,b)$ should obey inequality $|S| \leq 1$. In the case of pure imaginary elastic scattering amplitude  we obtain $S=1-2f$. The reflective scattering mode corresponds to negative values of the function $S$ \cite{refl}. 
The term reflective was adopted from optics.

 Evolution of central density of a scatterer with the energy and appearance of the reflective scattering mode  can be illustrated as the  scattering of  hadrons with hard cores (Fig. 1):
 \begin{figure}[hbt]
 	\hspace{-1cm}
 	\resizebox{16cm}{!}{\includegraphics{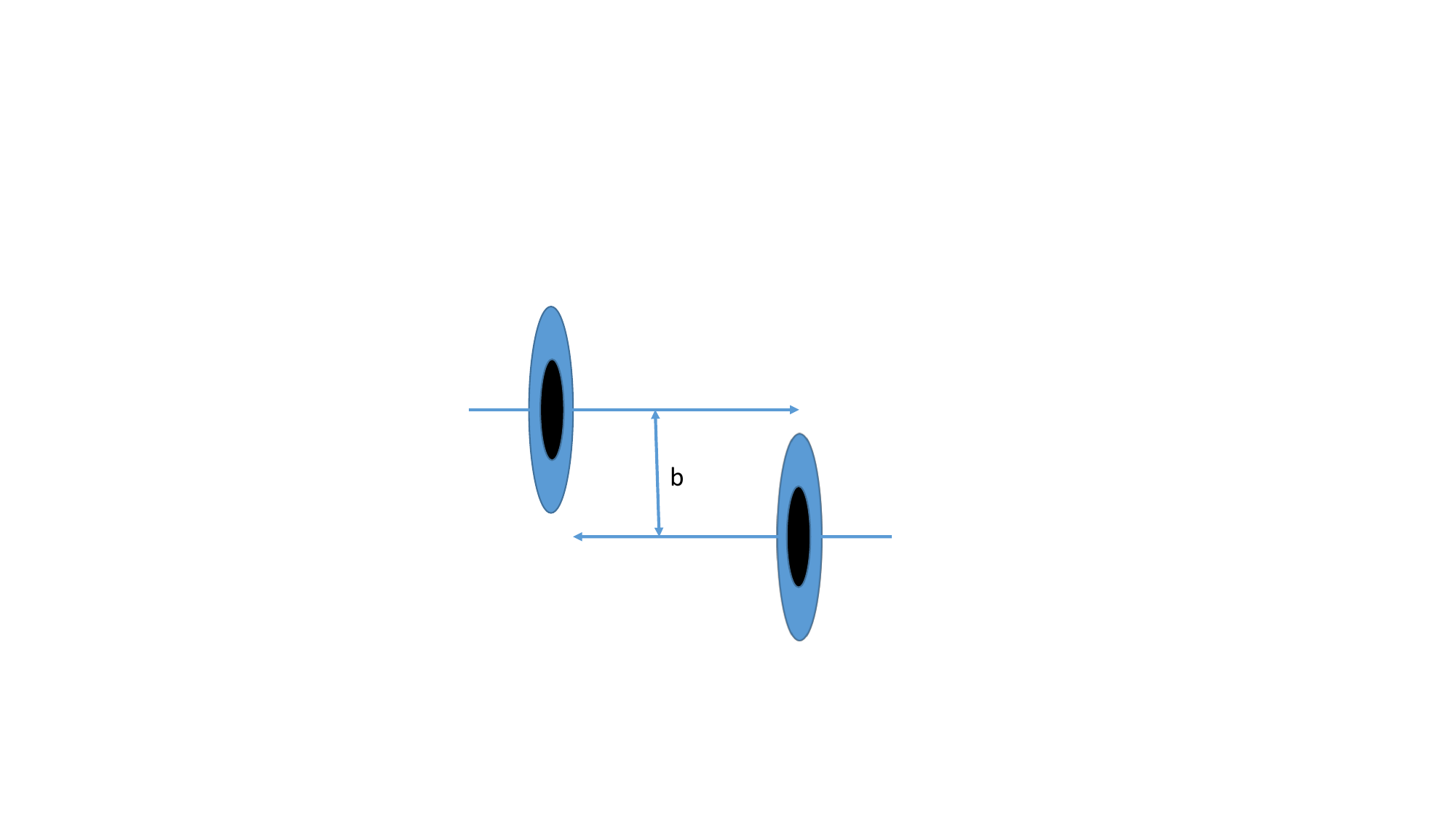}}		
 	\vspace{-2cm}
 	\caption{Illustration of hadron scattering  in the reflective scattering mode. It corresponds to an appearance of hadron repulsive cores   \cite{weis,pl21}.}	
 \end{figure}
 
This picture  assumes that the unitarity saturation takes place in the limit of $s\to\infty$, i.e the imaginary part of  partial elastic scattering amplitude ${\Im}f_l(s)$ tends to unity. 

In the impact parameter representation, the amplitude $f(s,b)$  obey the unitarity relation: 
\begin{equation} \label{unflb}
	\mbox{Im} f (s,b)=h_{el}(s,b)+ h_{inel}(s,b).
\end{equation}
The function $h_{ el}(s,b)\equiv |f(s,b)|^2$ is a contribution of the elastic intermediate state while $h_{inel}(s,b)$ is  a contribution of the inelastic intermediate states.

A central profile of the elastic overlap function and peripheral form of the inelastic overlap function are encoded into the relation between derivatives (in the case of a pure imaginary scattering amplitude) when the reflective scattering mode ($S<0$) occurs
\begin{equation}\label{cent}
	\frac {\partial h_{el}}{\partial s}=\left( \frac{1-S}{S}\right)\frac{\partial h_{inel}}{\partial s}.
\end{equation}
 The
qualitative $b$--dependencies of $h_{el}$ and $h_{inel}$ are presented in Fig. 2 where arrows indicate changes with the energy of the respective profiles. The  evolution in the opposite directions  of the elastic and inelastic overlap funcitions is a characteristic feature of the reflective scattering. It is to be noted  that the both functions change in the same way in the shadow scattering mode.
\begin{figure}[hbt]
	\vspace{-0.42cm}
	\hspace{-1cm}
	\resizebox{15cm}{!}{\includegraphics{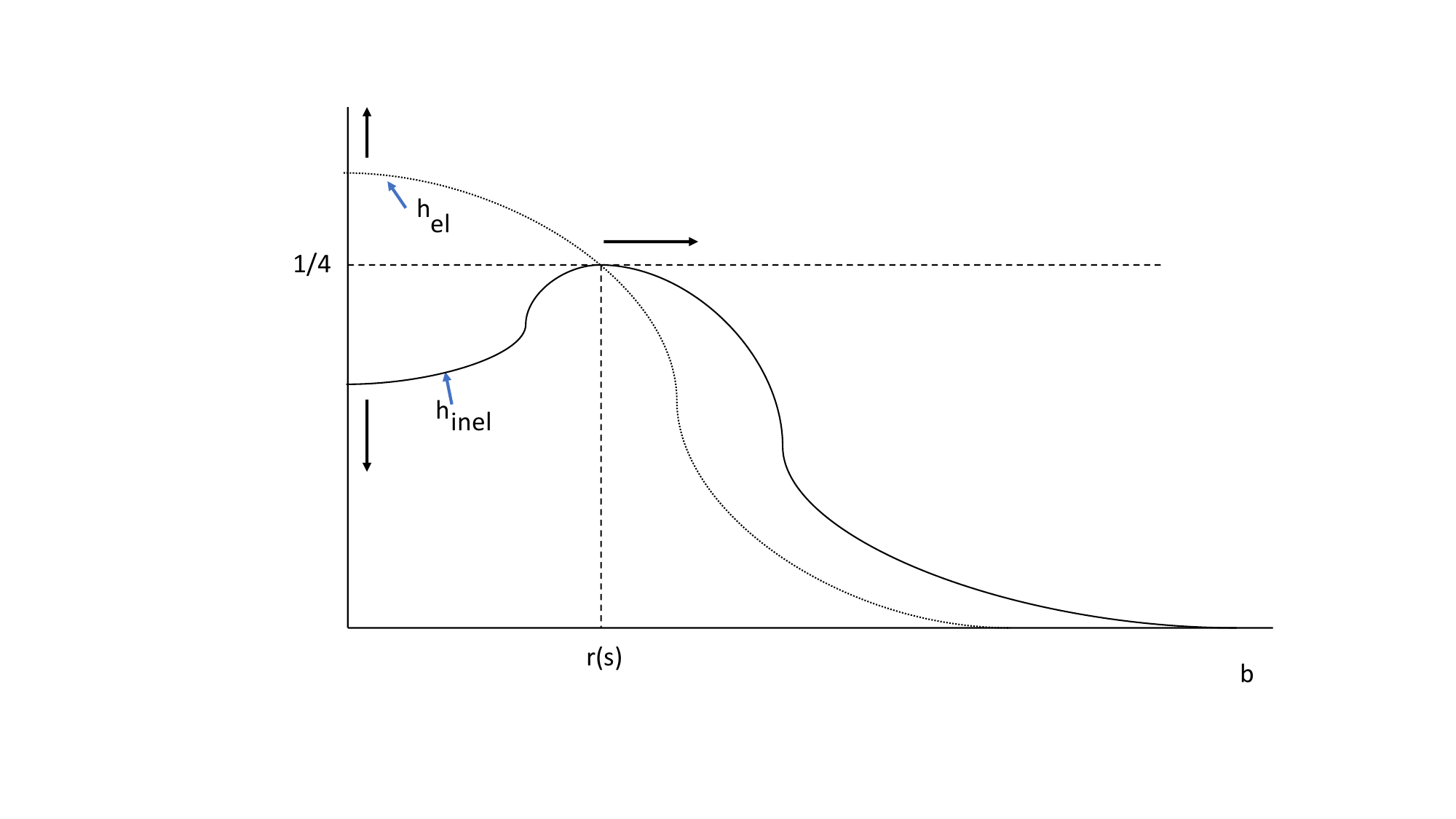}}		
	\vspace{-1cm}
	\caption{Profiles and energy evolution (indicated by arrows) of the elastic and inelastic overlap functions.}	
\end{figure}	 

The centrality definition in the case of hadron interactions is    \cite{cent}: 
\begin{equation}\label{cent}
	c^b(s)\equiv\frac{\sigma^b_{tot}(s)}{\sigma_{tot}(s)},
\end{equation} 
where 
\begin{equation}
\sigma^b_{tot}(s)=8\pi\int_0^b \mbox{Im} f(s,b')b'db'.	
\end{equation}
  Note, that $\sigma^b_{tot}(s)\to \sigma_{tot}(s)$ at $b\to\infty$. In. Eq. (\ref{cent}) the total cross--section replaces the inelastic one used in the case of nuclear scattering. The straightforward extension of the nuclear scattering case to hadron collisions would not lead to correct representation of the impact parameter.

The definition of centrality reflects the level of perepherality under hadron collisions. The origin can obviously be traced from the impact parameter concept but the appeal to centrality and other cumulative collision characteristics is dictated by the complexity of high energy dynamics.

The cumulative properties of the inelastic events have been discussed in \cite{inact}. 
The introduced quantity $a_{inel}^b(s)$ closely related to the inelastic overlap function  is a way to express a degree of peripherality for the inelastic events. 

Here we concentrate on cumulative characteristics of the elastic events and their relation to cumulative properties of the inelastic events and centrality related through unitarity.
\section{Cumulative  activity of elastic events  in hadron interactions}
Definition of cumulative elastic activity in hadron interactions  is similar to  definition of the inelastic activity \cite{inact}:
\begin{equation}\label{act}
a^b_{el}(s)\equiv\frac{\sigma^b_{el}(s)}{\sigma_{el}(s)},
\end{equation}
where 
\begin{equation}
\sigma^b_{el}(s)=2\pi \int_0^b\sigma_{el}(s,b')b'db'
\end{equation}
and  $\sigma_{el}(s,b)$ is the differential contribution ($\sigma_{el}(s,b)\equiv d\sigma_{el}/d^2\vb{b}$) to the elastic cross--section at the impact parameter $\vb{b}$  ($b\equiv|\vb{b}|$). 
The function $a^b_{el}(s)$ is positive, it  varies from zero to unity and determines the fraction of elastic events generated in hadron collisions with the impact parameter values in the interval  from zero to $b$. 

The dimensionless function $\sigma_{el}(s,b)$  is proportional to the elastic overlap function $h_{el}(s,b)$: $$\sigma_{el}(s,b)=4h_{el}(s,b).$$
Evidently, the differential elastic activity is determined by the elastic overlap function  $h_{el}(s,b)$:
\begin{equation}\label{pin}
\frac{\partial a^b_{el}(s)}{\partial b}=\frac{8\pi b}{\sigma_{el}(s)} h_{el}(s,b).
\end{equation}

 The advantage of using the cumulative  elastic activity is related to  a probabilistic nature of the impact parameter  reconstruction.
  The combined studies of centrality and cumulative elastic and inelastic activities in hadron interactions would be useful  for the  studies of hadron interaction dynamics. 
Relationship between these quantities \color{blue} directly \color{black} follows from unitarity: 
\begin{equation}\label{relc}
c^b(s)=a_{el}^b(s)\frac{\sigma_{el}(s)}{\sigma_{tot}(s)}+a_{inel}^b(s)\frac{\sigma_{inel}(s)}{\sigma_{tot}(s)}
\end{equation}
\color{blue} This relation allows  to calculate elastic events activity in hadron collisions from centrality and inelastic events activity without performing dedicated measurements of the observables associated with elastic scattering. Prospects of the experimental studies on the ground of Eq. (\ref{relc}) which is supposed to be valid at high but finite energies are discussed in Section 3.
\color{black}
\color{blue}

Eq. (\ref{relc}) \color{blue}  represents a central result completing  discussions of \cite{cent,inact}. It \color{black} has a form similar to the relation between  averaged values of the impact parameter $\langle b^2\rangle_{i}$ $(i=tot,el,inel)$   \cite{adj,web}. However, the interrelation between   the functions $\langle b^2\rangle_{i}$ is less sensitive to  the peculiarities of the impact parameter dependencies   since, contrary to Eq. (\ref{relc}), it includes quantities integrated over an entire region of the impact parameter variation.  

Eq. (\ref{relc}) \color{blue} can be considered as \color{black}  another expression of the unitarity relation projected onto the space of the relative event's numbers.
It is to be noted that there are no similar relations between centrality and events activities  for the cases of nucleus--nucleus  or proton--nucleus collisions \color{blue} since complete set of states is formed by the on--shell hadronic states. \color{black}

To  illustrate and discuss  energy dependence of the cumulative elastic  activity we use  the unitarization scheme based on representation of the scattering amplitude $f(s,b)$ in a rational form.  This scheme allows to cover  the entire range of  the elastic scattering amplitude variation allowed by unitarity \cite{umat}.
$S$--matrix element $S(s,b)$ of the elastic scattering is written as the Cayley transform
 which maps nonnegative real numbers (in the pure imaginary case $f\to if$) to the interval $ [-1, 1]$ of $S(s,b)$ variation\footnote{For the general case, this one-to-one transform  is   {\it a solution} (in a rational form) of Eq. (\ref{unflb}) provided that $\Im U\geq 0$;  it maps upper 
 	half-- plane of complex values of $U$ onto a unit circle in the complex plane of $S$  [both $U$  and $S$ $(S=1+2if)$ are complex functions in this case];  the point $S=0$ is a regular one,  it corresponds to {\it finite} values of the energy and impact parameter \cite{mpla23}.}:
 \begin{equation}
 S(s,b)={[1-U(s,b)]}/{[1+U(s,b)]}. \label{umi}
 \end{equation}
  The  function $U(s,b)$ is considered as an input amplitude and is an object for application of the  unitarization procedure. The choice  of this function in the form 
  \begin{equation}\label{usb}
  	U(s,b)=g(s)\omega(b),
  \end{equation}
  has been described in \cite{inact}.
  In Eq. (\ref{usb}) $g(s)\sim s^\lambda$ ($\lambda>0$) at large values of $s$.
  Such function $g(s)$ implies saturation of unitarity  for the scattering amplitude $f(s,b)$.
  The   function $\omega(b)\sim \exp({-\mu b})$ has been chosen to meet  analytical properties of the scattering amplitude.
 The use of  Eq. (\ref{usb})
 allows  to \color{blue} obtain the asymptotic dependence at large $s$ and $b$--fixed of the cumulative elastic activity: \color{black}
 \begin{equation}
 \label{asymp}
 a^b_{el}(s)\propto[1- {C}{s^{-\lambda}}]{\ln^{-2} s}, \, C>0.
 \end{equation}
 This dependence  results from   unitarity saturation assumed by Eqs. (\ref{umi}) and (\ref{usb}). 
 \color{blue}
 Saturation of unitarity  corresponds to the principle of maximal strength of strong interactions proposed long ago by Chew and Frautchi \cite{chew,chew1}.   It leads to energy decrease ($b$ is fixed) of the  inelastic activity   $a^b_{inel}(s)$  to zero \cite{inact} and respective increase of the normalized elastic activity  $a^b_{el}(s)$ to unity as well as evolution of hadron interaction region  to a black ring  with inner reflective core \cite{inact} and peripheral shadowing. Studies of those quantities could also provide an important information on the asymptotic regime of QCD in soft scattering domain. 
 
  The choice of  the function $U(s,b)$ in the form of Eq. (\ref{usb}) is based on the analyticity of scattering matrix and the principle of maximal strenghth of strong interactions mentioned above. To perform the numerical estimates, one needs to use a particular model parameterization. The value of $\lambda$ may be obtained from the energy dependence of the slope parameter \cite{epl}. The positive constant $C$ is a model--dependent one and not known apriori. In any case, the Eq. (\ref{asymp}) corresponds to the asymptotic  energies. 
 This region corresponds to the energy values around $\sqrt{s}\simeq 10^{10}$ TeV \cite{mpla23}.
 \color{black}

\section{Conclusion. Comments on the experimental measurements.}

The  event classification by multiplicity of the final state  seems to be appropriate for the cumulative inelastic activity estimation  but it is not a relevant tool for centrality reconstruction since  the elastic channel contribution is  almost neglected in that case because the average number of produced particles  is much greater than two at high energies. However, the total fraction of the elastic events is about $30$\% at the LHC energies \cite{totem}.
 
 The relevant observable for centrality determination can  be a sum of the transverse energies of the final state particles measured in the detector equiped with a hadron calorimeter.
 The ALICE, ATLAS and CMS  experiments \cite{cms} are, in principle, feasible for   solution of this task. The  respective method   was described in \cite{rog}. \color{blue}
 
 Eq. (\ref{relc}) being a consequence of the unitarity, centrality and the cumulative activities definitions allows the use of the data on the total transverse energy and multiplicity to extract the cumulative activity of elastic events
  $a^b_{el}(s)$. 

\color{black} The $b$-dependent differential quantities are responsible for a number of  important  conclusions such as  transition to a peripheral form of the inelastic overlap function contrary to a persisting central distribution of the elastic overlap function over the impact parameter. This allows to highlight the structure of the hadron interaction region  \cite{inact}.

   Unitarization procedure as a way to satisfy  unitarity constraints  serves also to mitigation of the assumed strong energy dependence of the input amplitude under transition to the final scattering  amplitude. 
   Strong  energy dependence  of the input contribution transforms into  the  observed effects of  soft interactions dynamics.

    The  reflective ability  of the  inner part of the interaction region increases with the energy \cite{mpla23} along with  similar increase of the cumulative activity of elastic events.

Further searches for  experimental manifestations of the reflective  mode  in the elastic and inelastic hadron interactions at the available energies is   an important issue.   The measurements of centrality combined with   measurements of the cumulative elastic and inelastic activities in $pp$-collisions  can be promising for that purpose. 

\end{document}